\documentclass{elsart}
\usepackage{epsf}
\usepackage[dvips]{graphicx}
\newcommand{\eg}{{\it e.g.}}
\newcommand{\ie}{{\it i.e.}}

\begin{document}
\begin{frontmatter}
\title{{\small\rm
\vspace{-.25in}\rightline{IIT-HEP-01/2}
}
Progress in Absorber R\&D for Muon Cooling\thanksref{talk}}
\thanks[talk]{Presented at the {\sl 
3rd International Workshop on Neutrino Factory Based on Muon Storage Rings 
(NuFACT'01)}, May 24--30, 2001, Tsukuba, Japan.}
\author{D. M. Kaplan, E. L. Black, M. Boghosian, K. W. Cassel, R. P. Johnson}
\address{Illinois Institute of Technology,
Chicago, IL 60616, USA}
\author{S. Geer, C. J. Johnstone, M. Popovic}
\address{Fermilab,
Batavia, IL 60510 USA}
\author{S. Ishimoto, K. Yoshimura}
\address{KEK,
Tsukuba-shi, Ibaraki-ken 305-0801, Japan}
\author{L. Bandura, M. A. Cummings, A. Dyshkant, D. Hedin, D. Kubik}
\address{Northern Illinois University,
DeKalb, IL 60115, USA}
\author{C. Darve}
\address{Northwestern University, Evanston, IL, USA}
\author{Y. Kuno}
\address{Osaka University,
Osaka 560-0043, Japan}
\author{D. Errede, M. Haney,  S. Majewski}
\address{University of Illinois at Urbana-Champaign,
Urbana, IL 61801, USA\\}\address{{\rm \large and}\\}
\author{M. Reep, D. Summers}
\address{University of Mississippi,
University, MS 38677, USA}
\begin{abstract}

A stored-muon-beam neutrino factory may require transverse ionization cooling of the muon beam.
We describe recent progress in research and development on energy
absorbers for muon-beam cooling carried out by a
collaboration of university and laboratory groups.

\end{abstract}
\end{frontmatter}

\section{Introduction}


To achieve the small emittance typically required for beam acceleration, a
stored-muon-beam neutrino factory may require transverse ionization cooling of
the muon beam~\cite{FS1,FS2}.\footnote{Alternative designs without cooling have
also been proposed~\protect\cite{FFAG}.} Such cooling can be accomplished by
passing the beam through energy-absorbing material and accelerating structures,
both embedded within a focusing magnetic lattice; the rate of change
$d\epsilon_n/ds$ of normalized transverse emittance  with path length is then
given approximately by~\cite{Neuffer2,Fernow}
\begin{equation}
\frac{d\epsilon_n}{ds}\ =\
-\frac{1}{(v/c)^2} \frac{dE_{\mu}}{ds}\ \frac{\epsilon_n}{E_{\mu}}\ +
\ \frac{1}{(v/c)^3} \frac{\beta (0.014)^2}{2E_{\mu}m_{\mu}\ L_R}\,,
\label{eq1}
 \end{equation} 
where muon energy $E_\mu$  is in GeV, $\beta$ is the transverse amplitude
function of the lattice evaluated at the location of the absorber, and $L_R$ is the
radiation length of the absorber medium. 

Simulations show that enough transverse cooling can be achieved to build a
high-performance neutrino factory~\cite{FS2,Overarching-report}.   For example,
neglecting Coulomb scattering  (\ie, ignoring the last term in Eq.~\ref{eq1}),
for typical parameter values (\eg\ $\epsilon_n\approx10\,$mm$\cdot$rad and
200\,MeV/$c$ muon momentum) and a 10--15\% packing fraction of absorber
within the cooling channel, the cooling rate of Eq.~\ref{eq1} implies
transverse emittance reduction by a factor $1/e$ in $\approx50\,$m, about 3\%
of the muon decay length. In practice, with $\beta\approx 20-50\,$cm, one does
a factor $\approx2$ worse than this because of scattering and other
effects~\cite{FS2}.

To minimize the effects of Coulomb scattering of the muons as they pass through the absorber, it has been proposed to use liquid hydrogen (LH$_2$) as the energy-absorbing medium~\cite{Status-Report}. Key issues in absorber R\&D include coping with the large heat deposition by the intense ($\sim 10^{14}/$s) muon beam\footnote{Palmer has suggested~\protect\cite{Palmer-NuFACT01} that muon intensities an order of magnitude higher than this can be achieved, compounding the engineering challenge.} and minimizing scattering in the absorber-vessel windows.
Specifications of absorbers for some representative
cases are given in Table~\ref{tab:abs}. Our absorber-window R\&D program is discussed in \cite{Kaplan-PAC2001} and \cite{Kaplan-NuFACT00}. Here we give an overview of absorber R\&D and a summary of recent progress.

\begin{table}
\begin{center}
\caption{Specifications of typical LH$_2$ absorbers (from the ``Neutrino Factory Feasibility Study II" report~\cite{FS2}).}
\label{tab:abs}
\begin{tabular}{l|ccccc}
\hline\hline
Absorber & Length & Radius & Number  & Power & Window\\
 & (cm) & (cm) & needed & (kW) & thickness ($\mu$m)\\
\hline
Minicooling & 175 & 30 & 2 & $\approx$5.5 & \\
SFOFO 1 & 35 & 18 & 16 & $\approx $0.3 & 360* \\
SFOFO 2 & 21 & 11 & 36 & $\approx$0.1 & 220* \\
\hline\hline
\end{tabular}
\end{center}
* Design parameter for 1.2-atm maximum pressure
\end{table}

\section{Absorber development}

The heat deposited in the hydrogen by the muon beam can exceed 100 watts per
absorber (Table~\ref{tab:abs}). LH$_2$ targets  using an external cooling
loop~\cite{Mark} have been successfully operated in such a heat-deposition
regime~\cite{SAMPLE}, but engineering the fluid flow is a
challenge~\cite{Kaplan-NuFACT00,E158-milestone}. We are developing prototypes
using two design approaches~\cite{Kaplan-NuFACT00}: a conventional,
``flow-through" (cooling-loop) design, and a new approach using internal heat
exchange, in which driven convection provides mixing and transverse
flow~\cite{Ishimoto-NuFACT01}.

\begin{figure}
\centerline{
{\includegraphics[bb=60 50 252 350,clip]{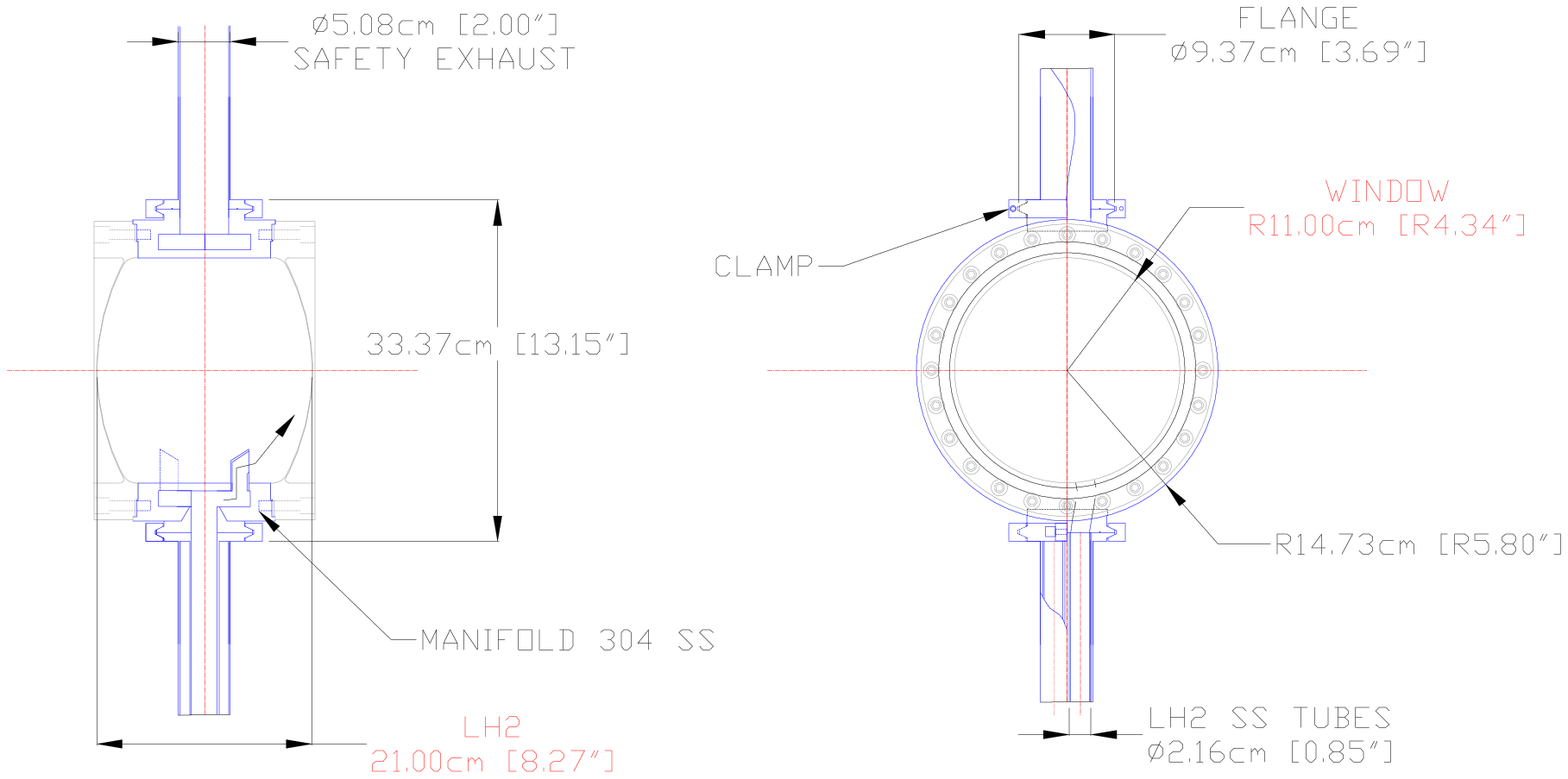}}
{\includegraphics[bb=335 50 540 350,clip]{ABSORBER2-1.eps}}}
\vspace{-2.65in}\hspace{2.15in}{\includegraphics[bb=270 220 307 235,clip]
{ABSORBER2-1.eps}}\\[1.58in]
\centerline{
\hspace{.1in}{\includegraphics[bb=260 95 275 110,clip]{ABSORBER2-1.eps}}}
\vspace{.5in}
\caption{Mechanical design of ``SFOFO 2" absorber (flow-through version).}
\label{fig:SFOFO}
\end{figure}

Fig.~\ref{fig:SFOFO} shows the mechanical layout of a flow-through absorber. Internal nozzles will be used to direct the fluid flow within the absorber to ensure adequate circulation and avoid dead zones or eddies. A room-temperature model with transparent plastic windows is under construction and will be used for first tests of the nozzle configuration using warm and cold water.

A critical-path item in absorber development is certification of safety by a Fermilab review committee. The stringent standards that must be met have been codified in \cite{LH2-safety} and include destructive testing of five windows of a given design before a sixth may be put into service. Pressure testing of a prototype window is underway~\cite{Kaplan-PAC2001}.

\section{Minicooling absorbers}

As shown in Table~\ref{tab:abs}, the Feasibilty Study II (FS2) neutrino factory
design~\cite{FS2} includes two large ``minicooling" absorbers. Their function
is to lower the muon energy from the optimal energy for capture by the 
channel's focusing optics to that which is
considered optimal for cooling.\footnote{Whether these energies are in fact the
optima has not yet been definitively established, but they are the
``provisional" optima assumed for the FS2 design.} At the same time they cool
the normalized transverse emittance by $\approx\sqrt{2}$.

In the FS2 design it is assumed that the minicooling absorbers are composed of LH$_2$. Such large LH$_2$ tanks with such high power dissipation go considerably beyond LH$_2$-target experience. However, the parameters are not dissimilar to those of the Fermilab 15-foot bubble chamber~\cite{15foot}. 

While LH$_2$ minicooling is surely technically feasible, we have
argued~\cite{FS2} that it is not necessarily the best choice.
Minicooling via \eg\ solid lithium or beryllium or liquid methane would also be
feasible and might well be preferable from an operational standpoint.  The
additional multiple scattering entailed with a higher-$Z$ absorbing material
could decrease the flux out of the neutrino factory. H. Kirk has simulated this
effect and has found the decrease to be only 5\% for lithium and 10\% for
beryllium~\cite{FS2}.  In principle this can easily be offset by raising the
solenoidal focusing field slightly, however, detailed design studies of this
idea remain to be carried out.

\section{Linac-area test facility}

To support absorber tests, a new experimental area is under construction at
Fermilab.  Its location near the end of the Linac makes available 201-
and 805-MHz power for high-power RF-cavity tests as well as 400-MeV $H^-$ beam
at high intensity. Table~\ref{tab:LTF} gives specifications of the beam. Our
planned program includes absorber bench tests and high-power 201-MHz RF-cavity
tests followed by assembly of an integrated prototype cooling cell (including
superconducting solenoids) for testing under radiation conditions typical for a
neutrino factory cooling channel. More generally, the Linac-area test facility
is a new experimental area at Fermilab available for any proposed experiment or
test that calls for 400-MeV $H^-$ or proton beam.

\begin{table}
\begin{center}
\caption{Linac-area test facility beam specifications.}
\label{tab:LTF}
\begin{tabular}{l|cc}
\hline\hline
Parameter & Minimum & Maximum \\
\hline
Beam Size ($\pm3\sigma$) at D.U.T.* (cm)&	1	& 30	 \\
Beam Divergence$^\dagger $ ($\pm3\sigma$) at D.U.T.* (mr)&	$\pm$0.5 &	$\pm$14	 \\
Number of Pulses per Second &	&	15 	\\
Number of Protons per Pulse ($10^{12}$)&	1.6	& 16      	 \\
Pulse Duration	($\mu$s)& 5.0	& 50	 \\
\hline\hline
\end{tabular}
\end{center}
* D.U.T. = Device Under Test\\
$^\dagger$ Min. divergence at max. size and vice versa.
\end{table}

\section{Gaseous absorbers}

A new idea has started to receive serious consideration: use of high-pressure
gaseous (rather than liquid) hydrogen as the energy-absorbing medium.  If the
gas is allowed to fill the entire cooling channel instead of being confined to
roughly 10\% of the channel length (as in current designs), matching the energy
loss to the RF accelerating gradient requires a factor $\sim10^2$ in density
compared to that at STP.  The pressure needed, especially if the hydrogen is
cooled to liquid-nitrogen temperature, is then comparable to what has been used
in the past for gaseous Cherenkov counters: about 20 atm.

Upon first consideration such an approach would appear to have significant
drawbacks.  These include the need for thick windows to withstand the pressure
as well as the introduction of material inside the RF cavities, which could
cause breakdown and (from Eq.~\ref{eq1}) degrade the cooling rate (via multiple
scattering at high-beta points of the lattice). However, 
calculations~\cite{mucool195} show that the cooling performance can in fact be
{\em enhanced} by use of gaseous absorbers: the many thin windows used in the
LH$_2$ case are replaced by only two thick windows, which degrade the final
emittance negligibly, and the dense gas  inside the cavities in fact {\em
suppresses} breakdown~\cite{Meeks}. The recent development of cooling lattices
with constant $\beta$~\cite{Monroe} alleviates the last of the drawbacks.
Further potential advantages include a more adiabatic cooling process, in which
the energy loss and acceleration occur continuously and muon momentum swings are
reduced, a slightly shorter overall channel length, which reduces muon decay losses,
and improvement of RF efficiency via the decrease of cavity resistivity at low
temperature.

A number of questions remain, including whether Paschen's Law (for high-voltage breakdown) is applicable in this regime of frequency, gas density, and radiation level, whether LN$_2$-temperature operation of 201-MHz RF cavities is indeed more economical when refrigeration costs are factored in, whether RF couplers can be designed to withstand 20-atm differential pressure, and whether constant-$\beta$ cooling channels (even with gaseous absorber) are cost-effective compared to other proposed approaches. These will be subjects of R\&D in the coming year.

\section{Acknowledgements}

We thank P. Lebrun and A. Tollestrup for useful discussions. This work was supported in part by
the U.S. Dept.\ of Energy, the National Science Foundation, Monbukagakusho (the
Ministry of Education, Culture, Sports, Science and Technology) of the
Goverment of Japan, the Illinois Board of Higher Education, and the Illinois
Dept.\ of Commerce and Community Affairs.


\begin{thebibliography}{99}

\bibitem{FS1}

``A Feasibility Study of a Neutrino Source Based on a Muon Storage Ring,"
N. Holtkamp and D. A. Finley, eds., FERMILAB-PUB-00-108-E (2000).

\bibitem{FS2}
``Feasibility Study-II of a Muon-Based Neutrino Source," S. Ozaki {\it et al.}, eds.,  June 14, 2001.

\bibitem{Neuffer2}
D. Neuffer, in {\bf Advanced Accelerator Concepts}, F. E. Mills, ed., AIP 
Conf.\ Proc.\ {\bf 156} (American Institute of Physics, New York, 1987), p.~201.

\bibitem{Fernow}
R. C. Fernow and J. C. Gallardo, Phys.\ Rev.\ E {\bf 52}, 1039 (1995).

\bibitem{Overarching-report}
``Summary Report on Neutrino Factory and Muon Collider," R. Raja {\it et al.}, eds., to be submitted to Phys.\ Rev.\ ST Accel.\ Beams.

\bibitem{Status-Report}
C. Ankenbrandt {\it et al.}, Phys.\ Rev.\ ST Accel.\ Beams {\bf 2}, 
081001, 1--73 (1999).

\bibitem{Kaplan-PAC2001}
D. M. Kaplan {\it et al.},  IIT-HEP-01/2, to appear in Proc.\ PAC2001.

\bibitem{Kaplan-NuFACT00}
D. M. Kaplan {\it et al.},  Proc.\ NuFACT00, IIT-HEP-00/1, physics/0008014 (to appear in Nucl.\ Instrum.\ Meth.)

\bibitem{FFAG}
S. Machida, this Workshop.


\bibitem{Palmer-NuFACT01}
R. B. Palmer, this Workshop.

\bibitem{Mark}
J. W. Mark, SLAC-PUB-3169 (1984) and references therein.

\bibitem{SAMPLE}
E. J. Beise {\it et al.}, Nucl.\ Instrum.\ Meth.\ {\bf A378} (1996) 383.


\bibitem{E158-milestone}
R. W. Carr {\it et al.}, SLAC-Proposal-E-158, July 1997,\\ and
``E158 Liquid Hydrogen Target Milestone Report,"
http://www.slac.stanford.edu/exp/e158/documents/target.ps.gz (April 21, 1999). 

\bibitem{Ishimoto-NuFACT01}
S. Ishimoto, this Workshop.

\bibitem{LH2-safety}
``Guidelines for the Design, Fabrication, Testing, Installation and Operation 
of Liquid Hydrogen Targets," Fermilab,
Rev.\ May 20, 1997; J. Kilmer, private communication.

\bibitem{15foot}
P.C.VanderArend {\it et al.}, {\bf 15-Foot Bubble Chamber Safety Report}, National Accelerator Laboratory Report NAL-48-A-2624, July1972, Vol.3.

\bibitem{mucool195} 
R. Johnson and D. M. Kaplan, MuCool Note 195, March 2001 (see
http://www-mucool.fnal.gov/notes/notes.html).

\bibitem{Meeks}
J. M. Meek and J. D. Craggs, {\bf Electrical Breakdown in Gases}, John Wiley \& Sons, 1978,
p. 557.

\bibitem{Monroe}
J. Monroe {\it et al.},  Phys.\ Rev.\ ST Accel.\ Beams {\bf 4}, 041301 (2001).



\end{thebibliography}
\end{document}